\documentclass[aps,reprint,superscriptaddress,amsmath,amssymb]{revtex4-2}

\usepackage{graphicx}
\usepackage{tikz-feynman}  
\usepackage{hyperref}  

\usepackage{slashed}
\usepackage[shortlabels]{enumitem}
\usepackage{mathtools}    
\usepackage{mathrsfs}     
\usepackage{bbm}          
\usepackage{slashed}      
\usepackage{braket}       
\usepackage{tensor}       

\usepackage{array}
\newcolumntype{C}{>{$}c<{$}}

\newcommand{\spdenom}[1]{\spa{1}{2}\spa{2}{3}\cdots\spa{n}{1}}
\newcommand{\spa}[2]{\left\langle#1\,#2\right\rangle}
\newcommand{\spb}[2]{\left[#1\,#2\right]}
\newcommand{\floor}[1]{\lfloor#1\rfloor}
\newcommand{\tr}{\text{tr}}
\newcommand{\Gr}{\text{Gr}}
\newcommand{\sig}{\sigma}
\newcommand{\lam}{\lambda}
\newcommand{\eps}{\epsilon}

\begin{document}


\title{Two-loop QCD Amplitudes from the Chiral Algebra Bootstrap}


\author{Anthony Morales}
\email{moralesa@uw.edu}
\affiliation{Department of Physics,
University of Washington, Seattle, Washington 98195}
\affiliation{SLAC National Accelerator Laboratory,
Stanford University, Stanford, CA 94309}
\affiliation{Department of Physics,
Stanford University, Stanford, CA 94305}


\date{\today}

\begin{abstract}
We show that the chiral algebra bootstrap, which computes form factors of twistorial theories, can help determine two-loop amplitudes in massless QCD. We give an $n$-gluon result for a previously unknown partial amplitude of the two-loop all-plus-helicity QCD amplitude by utilizing supersymmetry Ward identities and known chiral algebra bootstrap results. We then show that the full-color two-loop $n$-gluon amplitude of QCD with $n_f$ quark flavors can be obtained from certain two-loop form factors of twistorial theories and one-loop and tree-level amplitudes. Chiral algebra bootstrap results exist for these form factors when all gluons have positive helicities. Hence, the bootstrap simplifies the computation of these two-loop amplitudes by one loop level.
\end{abstract}


\maketitle

\section{\label{sec:intro}Introduction}

Over the last few decades, there has been dramatic progress in the computation of higher-point and higher-loop scattering amplitudes in quantum field theories. This endeavor has led to an improved understanding of the mathematical structure of many theories (see ref.~\cite{Elvang:2013cua} for a review), while also providing more precise predictions for physical observables \cite{Heinrich:2020ybq}.

Many of the advances in our understanding of amplitudes have come from theories with relatively nice analytical structures, such as $\mathcal{N}=4$ supersymmetric Yang-Mills (SYM) \cite{Arkani-Hamed:2022rwr,Caron-Huot:2020bkp,Caron-Huot:2019vjl,Dixon:2020cnr,Dixon:2023kop}. However, progress in non-supersymmetric gauge theories, such as QCD, has been much slower. The first two-loop results were given in refs.~\cite{Bern:2000dn,Anastasiou:2000kg,Anastasiou:2000ue,Anastasiou:2001sv,Glover:2001af,Bern:2002tk} for the scattering of four partons. There have been remarkable recent advances in computing the full-color all-helicity massless QCD amplitudes for the scattering of five partons at two loops~\cite{Agarwal:2023suw,DeLaurentis:2023nss,DeLaurentis:2023izi} and for the scattering of four partons at three loops~\cite{Caola:2021rqz,Caola:2021izf,Caola:2022dfa}. For the all-plus-helicity configuration (when all scattering partons are outgoing gluons with positive helicities), the full-color pure Yang-Mills (YM) amplitudes have been computed for the scattering of up to seven gluons \cite{Dalgleish:2020mof,Dalgleish:2024sey}, while $n$-gluon results and conjectures exist for certain partial amplitudes \cite{Dunbar:2016cxp,Dunbar:2020wdh,Dalgleish:2024sey}. Full-color QCD amplitudes have a rather intricate analytic structure, and pushing directly to one more loop or one more leg may be difficult.

A few years ago, Costello and Paquette proposed a novel bootstrap method to compute non-supersymmetric gauge theory amplitudes, dubbed the \textit{chiral algebra bootstrap} \cite{Costello:2022wso,Costello:2022upu}. It exploits the fact that YM theory can be obtained from a perturbation around the (anti-)self-dual sector. Positive- and negative-helicity states of self-dual Yang-Mills (sdYM) can be shown to be in one-to-one correspondence with the operators of a two-dimensional chiral algebra (CA), whose operator product expansions (OPEs) capture the holomorphic collinear limits of the field-theory states \cite{Costello:2022wso}. Correlation functions of the CA give form factors of the field theory, and certain sdYM form factors reproduce amplitudes of ordinary YM. The CA bootstrap computes these form factors by using the CA OPEs to constrain their functional forms. 

The main limitation of this bootstrap method is that it only computes loop-level YM amplitudes when special matter content is added to the theory. These theories contain fermions that transform in special representations of the gauge group and/or a scalar with a fourth-order kinetic term that couples to the topological term $\tr(F\wedge F)$ of the theory. Four-dimensional theories with such matter are called \textit{twistorial}. The question remains: can the CA bootstrap aid in the computation of amplitudes in non-twistorial theories, like QCD?

In this letter, we answer this question in the affirmative. Using supersymmetry Ward identities, we show that a CA bootstrap result of ref.~\cite{Costello:2023vyy} gives a hitherto unknown partial amplitude appearing in the trace-basis color decomposition of the two-loop $n$-gluon all-plus-helicity QCD amplitude. We then show that the full-color $n$-gluon QCD amplitude is given by lower-loop-level amplitudes and two-loop twistorial amplitudes, the latter of which are computable using the CA bootstrap for the all-plus-helicity configuration. The two-loop QCD amplitude is a sum of three gauge-invariant quantities with different partons running in the loops: the pure-gluon amplitude, the single-fermion-loop contribution, and the two-fermion-loop contribution. Determining all three contributions to the QCD amplitude will require using three twistorial theories, whose properties we now explain.

\section{\label{sec:twisttheories}Twistorial Theories}
It has been shown that positive- and negative-helicity states of sdYM on twistor space are in one-to-one correspondence with local operators in a CA supported on a Riemann sphere \cite{Costello:2022wso}. The conformal blocks of the algebra correspond to local operators of the field theory; so, correlation functions in the CA give form factors of sdYM. Moreover, the OPEs in the algebra correspond to holomorphic collinear limits of the gluonic states. This equivalence suggests that the analytic properties of the chiral algebra OPEs can be used to ``bootstrap'' sdYM form factors. Indeed, this \textit{chiral algebra bootstrap} has been employed to compute form factors that give one- and two-loop amplitudes in ordinary $SU(N_c)$ gauge theories \cite{Costello:2022upu,Costello:2022wso,Costello:2023vyy,Dixon:2024mzh}, and it can compute, in principle, other form factors to higher loop orders using the results of ref.~\cite{Fernandez:2024qnu}.

A requirement for the existence of a CA is the associativity of its OPEs. Associativity fails at the first loop correction for pure sdYM. This failure is traced back to the fact that the one-loop all-plus amplitude is anomalous in the 6-dimensional twistor space uplift of the theory. The twistor anomaly can be removed via a Green-Schwarz mechanism \cite{Green:1984sg} by including a fourth-order scalar field that couples to the Yang-Mills topological term $\tr(F\smash{\widetilde{F}})$ \cite{Costello:2021bah,Costello:2022upu,Costello:2022wso}. However, this quartic ``axion'' can only remove double-trace contributions to the all-plus amplitude. Thus, this anomaly-cancellation mechanism requires that the gauge group have a quartic Casimir structure proportional to its quadratic one. Alternatively, the anomaly can be cured by introducing Weyl fermions living in a real representation whose quartic Casimir equals the adjoint representation's \cite{Costello:2023vyy}. A combination of the axion and fermions can also be employed simultaneously. A 4-dimensional theory with an anomaly-free twistor space description is called \textit{twistorial}.

The full anomaly-free sdYM Lagrangian on Minkowski space is given by
\begin{equation}
	\label{Lagrangian}
	\tr(BF_+) + \bar{\psi}\slashed{D}_R\psi
	-\frac{1}{2}(\partial^2\rho)^2 + \frac{g^2\lam_{G,R}}{8\pi\sqrt{3}}\rho~\tr(F\smash{\widetilde{F}}) \,,
\end{equation}
where $F_+$ is the self-dual component of the field strength $F$, $B$ is a self-dual two-form Lagrange multiplier field, $\widetilde{F}^{\mu\nu}:=\tfrac{i}{2}\eps^{\mu\nu\rho\sig}F_{\rho\sig}$ is the dual of $F$, $\psi$ is a Weyl fermion, $\rho$ is a real scalar field, and $g$ is the gauge coupling. The real matter representation $R$ and the value of the constant $\lam_{G,R}$ must satisfy the anomaly-free condition
\begin{equation}
	\label{anomfreecond}
	\tr_G(X^4)-\tr_R(X^4)=\lam_{G,R}^2\,\tr(X^2)^2\,,
\end{equation}
where $X$ is any Lie algebra element. The trace without a subscript is taken over the defining representation of the gauge group, and $\tr_G$ is the trace over the adjoint representation.

Twistorial theories have a trivial $S$-matrix on flat spacetimes \cite{Costello:2021bah,Costello:2022wso}. However, they have non-vanishing form factors, which are amplitudes in the presence of local operators. Form factors with multiple insertions of $\tfrac{1}{2}\tr(B^2)$ are of particular interest, as they reproduce amplitudes of non-self-dual gauge theory after sending the operator momenta to zero. This fact is seen by adding the term $\tfrac{1}{2}\tr(B^2)$ to the Lagrangian \eqref{Lagrangian} and integrating out $B$, yielding the YM Lagrangian plus a theta term. Form factors with a single insertion of $\tfrac{1}{2}\tr(B^2)$ reproduce two-minus- (MHV), one-minus-, and all-plus-helicity amplitudes in ordinary YM at tree, one-loop, and two-loop level, respectively, with more insertions giving higher-loop and more generic helicity amplitudes. Even though twistorial theories have no amplitudes, we will refer to the $\tfrac{1}{2}\tr(B^2)$ form factors with zero operator momenta as amplitudes, by slight abuse of terminology.

Because we will relate amplitudes of twistorial theories to those of QCD, we restrict ourselves to twistorial theories that have gauge group $SU(N_c)$ and with Weyl fermions transforming under $R=n_f\,(F\oplus\bar{F})$, with the values of $N_c$, $n_f$, and $\lam_{N_c,n_f}^2\equiv\lam_{SU(N_c),n_f\,(F\oplus\bar{F})}$ tuned to satisfy the anomaly-free condition \eqref{anomfreecond}. There are three classes of such twistorial theories, and they are defined in tab.~\ref{tab:theories} by the different possible choices for these three numbers.

\begin{table}
    \centering
    \begin{tabular}{|C|C|C|C|}
    \hline
        \text{Theory} & N_c & n_f & \lam^2_{N_c,n_f}\\
        \hline
        \text{I} & 2,3 & 0 & N_c+6 \\
        \text{II} & 2,3 & N_c+6 & 0 \\
        \text{III} & \text{arbitrary} & N_c & 6 \\
        \hline
    \end{tabular}
    \caption{This table provides the specific values of $N_c$, $n_f$, and $\lam^2_{N_c,n_f}$ that define the three classes of twistorial theories that we use. The massless Dirac fermions transform in $n_f$ copies of the fundamental representation of $SU(N_c)$. These numbers allow for the anomaly-free condition \eqref{anomfreecond} to be satisfied. }
    \label{tab:theories}
\end{table}

The CA bootstrap can compute the two-loop all-plus-helicity gluon amplitude for the three twistorial theories in tab.~\ref{tab:theories} \cite{Costello:2022upu,Costello:2023vyy,Dixon:2024mzh}. For $SU(2)$ and $SU(3)$, these amplitudes are given by two-loop all-plus QCD amplitudes with specific values for the number of quark flavors $n_f$, as well as lower-loop amplitudes involving internal axion propagators. In the next section, we review the various components of two-loop gluon amplitudes in QCD, in preparation for relating them to twistorial amplitudes.

\section{\label{sec:trbasis}Components of two-loop gluon amplitudes in QCD}

Two-loop gluon amplitudes in QCD can be decomposed into three gauge-invariant components distinguished by the type of particles running in the loops:
\begin{equation}
\label{QCDdecomp}
\mathcal{A}^{(2)}_\text{QCD} = \mathcal{A}^{(2)}_{G} + n_f\,\mathcal{A}^{(2)}_{F} + n_f^2\,\mathcal{A}^{(2)}_{F^2}
\,.
\end{equation}
The first term of the above decomposition, $\mathcal{A}^{(2)}_{G}$, is the pure-glue (YM) amplitude, in which all the particles in the loops are gluons. The second term, $\mathcal{A}^{(2)}_{F}$, is the sum of all Feynman diagrams that contain a single fermion loop, and the third term, $\mathcal{A}^{(2)}_{F^2}$, has two fermion loops.

Amplitudes in a non-abelian gauge theory can be decomposed into a basis of color traces taken over the fundamental (defining) representation of the Lie algebra of the gauge group. The color decomposition for $SU(N_c)$ gauge theory with $n_f$ quark flavors is
\begin{multline}
\mathcal{A}_n^{(2)} =\sum_{b=1}^{\floor{n/3}+1}
\sum_{c=2b-1}^{\floor{(n+b+1)/2}}\sum_{\sig\in S_n/S_{n;b;c}}
\Gr_{n;b;c}(\sig)
\\
\times A_{n;b;c}(\sig)
\,,
\end{multline}
where
\begin{multline}
    \Gr_{n;b;c}(1,2,\dotsc,n) = \tr(t^{a_1}\cdots t^{a_{b-1}})
    \\
    \times \tr(t^{a_b}\cdots t^{a_{c-1}})
    \,\tr(t^{a_c}\cdots t^{a_n})
\end{multline}
with
\begin{align}
    \Gr_{n;1;1}(1,2,\dotsc,n) &= N_c^2\,\tr(t^{a_1}\cdots t^{a_n})
    \\
    \Gr_{n;1;c}(1,2,\dotsc,n) &= N_c\,\tr(t^{a_1}\cdots t^{a_{c-1}})\,\tr(t^{a_c}\cdots t^{a_n})
    \,,
\end{align}
and $S_{n;b;c}$ is the set of all permutations that leave $\Gr_{n;b;c}$ invariant.

\begin{figure}
    \centering
    \begin{tikzpicture}[scale=0.5]
        \begin{feynman}
            \vertex (v1) at (-2,-1);
            \vertex (v2) at (-2,1);
            \vertex (v3) at (2,1);
            \vertex (v4) at (2,-1);
            \vertex (v5) at (0,1);
            \vertex (v6) at (0,-1);
            \vertex (g6) at (0,-2) {\(P_1\)};
            \diagram*{
                (v1) -- [anti fermion] (v6)
                -- [gluon] (v4) 
                -- [gluon] (v3) 
                -- [gluon] (v5)
                -- [anti fermion, swap] (v2) 
                -- [anti fermion] (v1),
                (v5) -- [fermion, color=red] (v6)
                -- [opacity=0] (g6),
            };
        \end{feynman}
    \end{tikzpicture}
    \begin{tikzpicture}[scale=0.5]
        \begin{feynman}
            \vertex (v1) at (-1.75,0);
            \vertex (v2) at (1.75,0);
            \vertex (v3) at (-0.5,0);
            \vertex (v4) at (0.5,0);
            \vertex (g) at (0,-2) {\(P_2\)};
            \diagram*{
                (v3) -- [gluon] (v1),
                (v2) -- [gluon] (v4),
                (v4) -- [dotted] (v3),
                (v2) -- [opacity=0] (g),
            };
            \draw[gluon] (-2.75,0) circle [radius=1];	
            \draw[anti fermion, rotate=180] (-2.75,0) circle [radius=1];		
        \end{feynman}
    \end{tikzpicture}
    \begin{tikzpicture}[scale=0.5]
        \begin{feynman}
            \vertex (v1) at (-2,-1);
            \vertex (v2) at (-2,1);
            \vertex (v3) at (2,1);
            \vertex (v4) at (2,-1);
            \vertex (v5) at (0,1);
            \vertex (v6) at (0,-1);
            \vertex (g6) at (0,-2) {\(P_3\)};
            \diagram*{
                (v1) -- [anti fermion] (v6)
                -- [anti fermion] (v4) 
                -- [anti fermion] (v3) 
                -- [anti fermion, swap] (v5)
                -- [anti fermion, swap] (v2) 
                -- [anti fermion] (v1),
                (v5) -- [gluon, color=red] (v6)
                -- [opacity=0] (g6),
            };
        \end{feynman}
    \end{tikzpicture}
    \begin{tikzpicture}[scale=0.5]
        \begin{feynman}
            \vertex (v1) at (-1.75,0);
            \vertex (v2) at (1.75,0);
            \vertex (v3) at (-0.5,0);
            \vertex (v4) at (0.5,0);
            \vertex (g) at (0,-2) {\(P_4\)};
            \diagram*{
                (v3) -- [gluon] (v1),
                (v2) -- [gluon] (v4),
                (v4) -- [dotted] (v3),
                (v2) -- [opacity=0] (g),
            };
            \draw[anti fermion] (-2.75,0) circle [radius=1];	
            \draw[anti fermion, rotate=180] (-2.75,0) circle [radius=1];			
        \end{feynman}
    \end{tikzpicture}
    \caption{Two-loop topologies of diagrams contributing to $\mathcal{A}^{(2)}_{F}$ and $\mathcal{A}^{(2)}_{F^2}$. Planar diagrams are given by attaching tree graphs to any black line at a single leg. Non-planar diagrams are those with topologies $P_1$ and $P_3$ with trees that attach to the red propagator shared by the two loops. The partial amplitudes $A_n^{(1,1)}$, $A_n^{(-1,1)}$, and $A_n^{(0,2)}$ are completely built from the single-trace terms of planar graphs with topologies $P_1$ \& $P_2$, $P_3$, and $P_4$, respectively.}
    \label{fig:tops}
\end{figure}
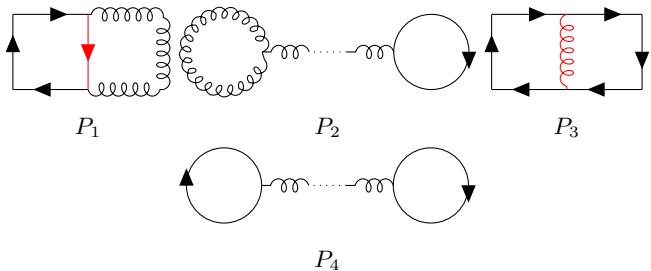

The partial amplitudes $A_{n;b;c}$ can be expanded in terms of powers of $N_c$ and $n_f$. For the single-trace partial amplitudes, this expansion is
\begin{multline}
\label{partamps}
    A_{n;1;1} = A_n^{(2,0)}+\frac{1}{N_c^2}\,A_n^{(0,0)}
    \\
    + \frac{n_f}{N_c}\,A_n^{(1,1)}
    + \frac{n_f}{N_c^3}\,A_n^{(-1,1)}
    +\frac{n_f^2}{N_c^2}\,A_n^{(0,2)}
    \,.
\end{multline}
The partial amplitudes $A_n^{(2,0)}$, $A_n^{(1,1)}$, $A_n^{(-1,1)}$, and $A_n^{(0,2)}$ are solely built from the single-trace terms of planar Feynman diagrams, whereas $A_n^{(0,0)}$ only has non-planar contributions. The topologies for the amplitudes with fermion loops are shown in fig.~\ref{fig:tops}. Even though $A_n^{(-1,1)}$ is subleading in powers of $N_c$, it only receives contributions from planar diagrams with the $P_3$ topology. We prove this statement in the supplemental materials.

\section{\label{sec:Am11}An $n$-gluon partial amplitude result}
We now show that the leading-color result for the two-loop all-plus $n$-gluon amplitude of theory III gives the partial amplitude $A_n^{(-1,1);\,\text{all-plus}}$. The argument requires the supersymmetry Ward identity
\begin{equation}
    \label{SWI}
    \mathcal{A}_{\mathcal{N}=1~\text{SYM}}^\text{all-plus}=0
    \,.
\end{equation}
The vanishing of the all-plus amplitude in $\mathcal{N}=1$ SYM happens order by order in the loop expansion. Recall that $\mathcal{N}=1$ SYM is the supersymmetric $SU(N_c)$ gauge theory with massless Weyl fermions transforming in the adjoint representation. 

Amplitudes of $\mathcal{N}=1$ SYM are comprised of the same diagrams that give the QCD amplitude but with the fermion color matrices $t_{n_f\,(F\oplus\bar{F})}^a$ replaced with adjoint representation matrices $t_G^a$. Thus, all planar diagrams with fermion loops contribute to the leading-color single-trace terms of the $\mathcal{N}=1$ SYM amplitude. At two loops, the leading-color partial amplitude of $\mathcal{N}=1$ SYM in terms of the partial amplitudes \eqref{partamps} is
\begin{equation}
\label{N=1partamp}
 A^{(2,0)}_{\mathcal{N}=1;n}=A_n^{(2,0)}
    + A_n^{(1,1)}
    +A_n^{(0,2)}
    - A_n^{(-1,1)}
    \,,
\end{equation}
where the minus sign in front of $A_n^{(-1,1)}$ follows from the fact that the single-trace contribution from a diagram with the $P_3$ topology is $-\frac{n_f}{N_c}\,\tr(\dots)$, whereas it is $+N_c^2\,\tr(\dots)$ when the fermions live in the adjoint. When restricted to the all-plus-helicity configuration, the supersymmetric partial amplitude \eqref{N=1partamp} vanishes
\begin{equation}
    \label{SWIpartamp}
    A^{(2,0);\,\text{all-plus}}_{\mathcal{N}=1;n}=0
    \,.
\end{equation}

Now consider the leading-color partial amplitude of twistorial theory III. This amplitude is given by setting $n_f=N_c$ in the QCD amplitude \eqref{partamps} and extracting the $\mathcal{O}(N_c^2)$ term
\begin{equation}
    \label{partampIII}
    A^{(2,0)}_{\text{III};n}=A_n^{(2,0)} + A_n^{(1,1)} + A_n^{(0,2)}
    \,.
\end{equation}
This partial amplitude does not receive contributions from diagrams involving the internal exchanges of the axion; single-trace axion contributions occur at $\mathcal{O}(N_c^0)$. Restricting to the all-plus amplitude and using eqs.~\eqref{N=1partamp} and \eqref{SWIpartamp}, the leading-color partial amplitude of theory III \eqref{partampIII} reduces to
\begin{equation}
    A^{(2,0);\,\text{all-plus}}_{\text{III};n} = A_n^{(-1,1);\,\text{all-plus}}
    \,.
\end{equation}

The $n$-gluon all-plus single-trace partial amplitudes of theory III were computed with the CA bootstrap in ref.~\cite{Costello:2023vyy}. Using their result for $A^{(2,0);\,\text{all-plus}}_{\text{III};n}$, we get an $n$-gluon result for $A_n^{(-1,1);\,\text{all-plus}}$
\begin{widetext}
\begin{equation}
\label{partampres}
    A_n^{(-1,1);\,\text{all-plus}}(1,2,\dotsc,n) = \frac{1}{2(4\pi)^4}
    \sum_{1\leq i<j<k<l\leq n}\frac{\spb{i}{j}\spa{j}{k}\spb{k}{l}\spa{l}{i} + \spa{i}{j}\spb{j}{k}\spa{k}{l}\spb{l}{i}}{\spdenom{n}}
    \,.
\end{equation}
\end{widetext}
This expression, which uses standard spinor-helicity notation, is recognized to be proportional to the so-called ``even'' part of the one-loop all-plus amplitude \cite{Bern:1993qk,Mahlon:1993si}, named as such because the numerator of this expression is even under a parity transformation of the outgoing gluon momenta. We have checked eq.~\eqref{partampres} for $n=4$ and $n=5$ using the results of refs.~\cite{Bern:2000dn,Bern:2002zk} and \cite{Agarwal:2023suw}, respectively \footnote{The author is grateful to Federica Devoto for sharing with him the bare, unrenormalized 5-gluon amplitude.}.

The result \eqref{partampres} is consistent with four-dimensional unitarity cuts. Unitarity cuts of $A_n^{(-1,1);\,\text{all-plus}}$ always result in a product involving tree-level amplitudes with a quark-anti-quark pair and gluons of positive helicity only; these tree amplitudes are zero. Thus, $A_n^{(-1,1);\,\text{all-plus}}$ must be a rational function in the kinematic variables.

The expected factorization properties provide further evidence to support the validity of eq.~\eqref{partampres}. Two-loop collinear limits dictate that \cite{Bern:2004cz}
\begin{multline}
\label{collin}
    A_n^{(-1,1);\,\text{all-plus}}(\dotsc,a,a+1,\dotsc) \longrightarrow 
    \\
    \frac{1}{\sqrt{z(1-z)}\spa{a,}{a+1}}A_{n-1}^{(-1,1);\,\text{all-plus}}(\dotsc,P,\dotsc)
    \,,
\end{multline}
where the momenta of the two color-adjacent gluons are taken to $p_a\to zP$ and $p_{a+1}\to (1-z)P$, with $P$ being a light-like four-momentum and $z$ a number. A simple computation shows that the partial amplitude \eqref{partampres} satisfies the collinear limit \eqref{collin} \cite{Bern:1993qk}. Also, $A_n^{(-1,1);\,\text{all-plus}}$ lacks multi-particle factorization poles at $P_{i,j}^2=0$, where $P_{i,j}=p_i+p_{i+1}+\dots+p_j$ for $1\leq i<j\leq n$ and $i+2\leq j$, as expected for this partial amplitude with this helicity configuration.

The expression \eqref{partampres} could have been guessed based on these factorization properties and knowing the 4-point and 5-point amplitudes, similar to how the one-loop all-plus amplitude was originally computed in ref.~\cite{Bern:1993qk}. However, one often does not have control over the terms that are regular in these on-shell factorization limits. In contrast, the symmetries of the CA can determine the terms regular in OPE limits \cite{Costello:2022upu,Costello:2022wso,Costello:2023vyy,Dixon:2024mzh}.

It is interesting that the ``odd'' part of the one-loop all-plus amplitudes is missing in our expression for $A_n^{(-1,1);\,\text{all-plus}}$, but it appears in a conjecture for the partial amplitude $A_n^{(0,0);\,\text{all-plus}}$ \cite{Dunbar:2020wdh}. Moreover, the partial amplitude $A_n^{(0,0);\,\text{all-plus}}$ is missing the even part. Perhaps one can interpret these facts as a consequence of the supersymmetry Ward identity \eqref{SWI} and the amplitudes' relation to form factors of twistorial theories. 

\section{From twistorial theories to QCD}
\label{sec:QCDfromtwist}
Significant effort has been made to compute two-loop QCD amplitudes, with the state-of-the-art being all 5-parton QCD amplitudes \cite{Agarwal:2023suw, DeLaurentis:2023nss,DeLaurentis:2023izi} and the 7-point all-plus pure YM amplitude \cite{Dalgleish:2024sey}. The goal of this section is to show that the full-color $n$-gluon QCD amplitude can be obtained from two-loop amplitudes of the twistorial theories and one-loop and tree amplitudes involving the axion. When restricting to the all-plus-helicity configuration, two-loop twistorial amplitudes are computable using the CA bootstrap \cite{Costello:2022upu,Costello:2023vyy, Dixon:2024mzh}, thus, reducing the computational difficulty of the QCD amplitude from two loops to one loop.

The two-loop gluon amplitude in the theory of QCD plus the axion is
\begin{equation}
	\label{ampQCDax}
	\mathcal{A}^{(2)}_\text{QCD+ax}
	=
	\mathcal{A}^{(2)}_\text{QCD}
	+\lam_{N_c,n_f}^2\,\mathcal{A}^{(1)}_\text{ax}
	+\left(\lam_{N_c,n_f}^2\right)^2\,\mathcal{A}^{(0)}_{\text{ax}^2}
	\,.
\end{equation}
The two-loop amplitudes of the three twistorial theories are obtained from eq.~\eqref{ampQCDax} by setting $n_f$ and $\lam_{N_c,n_f}^2$ to the specific values given in tab.~\ref{tab:theories}. The amplitude \eqref{ampQCDax} receives contributions from one-loop amplitudes $\mathcal{A}^{(1)}_\text{ax}$ built from diagrams with a single axion propagator (e.g.~fig.~\ref{fig:loopax}) and from tree amplitudes $\mathcal{A}^{(0)}_{\text{ax}^2}$ with two axion propagators (e.g.~fig.~\ref{fig:treeax}). The one-loop axion amplitude is given by a sum of three gauge-invariant terms
\begin{equation}
    \label{1loopaxamp}
    \mathcal{A}^{(1)}_\text{ax} = \mathcal{A}^{[g]}_\text{ax}
	+n_f\,\mathcal{A}^{[f]}_\text{ax}
	+\mathcal{A}^{[g\to ax]}_\text{ax},
\end{equation}
where the superscripts denote the particle circulating in the loop. Gluons and fermions receive the superscripts $[g]$ and $[f]$, respectively. The superscript $[g\to ax]$ denotes that the axion is one of the propagators in the loop, with the remaining particles in the loop being gluons.

\begin{figure}
	\centering
    \begin{tikzpicture}[baseline={(0,0)},scale=0.5,rotate=180]
        \begin{feynman}
            \vertex (g1) at (0,-2);
            \vertex (g2) at (0,2);
            \vertex (g3) at (4,2);
            \vertex (g4) at (4,-2);
            \vertex (v1) at (1,-1);
            \vertex (v2) at (1,1);
            \vertex (v3) at (3,1);
            \vertex (v4) at (3,-1);
            \diagram*{
                (g1) -- [gluon] (v1),
                (v2) -- [gluon] (g2),
                (g3) -- [gluon] (v3),
                (v4) -- [gluon] (g4),
                (v1) -- [quarter right, gluon] (v4) -- [scalar] (v3) 
                -- [quarter right, gluon] (v2) -- [quarter right, gluon] (v1),
            };
        \end{feynman}
    \end{tikzpicture}
    \begin{tikzpicture}[baseline={(0,0)},scale=0.5]
        \begin{feynman}
            \vertex (g1) at (-3,-2);
            \vertex (g2) at (-3,2);
            \vertex (v1) at (-1.5,0);
            \vertex (v2) at (0.25,0);
            \coordinate (C) at (1.25,0);
            \def\R{1}
            \vertex (v3) at (1.25+0.5,0.866);
            \vertex (v4) at (1.25+0.5,-0.866);
            \vertex (v5) at (1.25+1,0);
            \vertex (g3) at (1.25+1,2);
            \vertex (g4) at (1.25+1,-2);
            \vertex (g5) at (3.5,0);
            \diagram*{
                (g1) -- [gluon] (v1) -- [gluon] (g2),
                (v1) -- [scalar] (v2),
                (g3) -- [gluon] (v3),
                (v4) -- [gluon] (g4),
                (v5) -- [gluon] (g5),
            };
            \draw[gluon] (C) ++(60:\R) arc (60:180:\R);
            \draw[gluon] (C) ++(0:\R) arc (0:60:\R);
            \draw[gluon] (C) ++(-60:\R) arc (-60:0:\R);
            \draw[gluon] (C) ++(180:\R) arc (180:300:\R);
        \end{feynman}
    \end{tikzpicture}
    \begin{tikzpicture}[baseline={(0,0)},scale=0.5]
        \begin{feynman}
            \vertex (g1) at (-3,-2);
            \vertex (g2) at (-3,2);
            \vertex (v1) at (-1.5,0);
            \vertex (v2) at (-0.5,0);
            \coordinate (C) at (1.75,0);
            \def\R{1}
            \vertex (v3) at (1.75+0.5,0.866);
            \vertex (v4) at (1.75+0.5,-0.866);
            \vertex (v5) at (1.75-1,0);
            \vertex (g3) at (1.75+1,2);
            \vertex (g4) at (1.75+1,-2);
            \vertex (g5) at (-0.5,-2);
            \diagram*{
                (g1) -- [gluon] (v1) -- [gluon] (g2),
                (v1) -- [scalar] (v2),
                (g3) -- [gluon] (v3),
                (v4) -- [gluon] (g4),
                (v5) -- [gluon] (v2),
                (v2) -- [gluon] (g5),
            };
            \draw[anti fermion] (C) ++(60:\R) arc (60:180:\R);
            \draw[anti fermion] (C) ++(-60:\R) arc (-60:60:\R);
            \draw[anti fermion] (C) ++(180:\R) arc (180:300:\R);
        \end{feynman}
    \end{tikzpicture}
	\caption{Examples of Feynman diagrams that contribute to the one-loop amplitude with a single axion propagator $\mathcal{A}^{(1)}_\text{ax}$. The left, middle, and right diagrams contribute to the amplitudes $\mathcal{A}^{[g\to ax]}_\text{ax}$, $\mathcal{A}^{[g]}_\text{ax}$, and $\mathcal{A}^{[f]}_\text{ax}$, respectively.}
	\label{fig:loopax}
\end{figure}
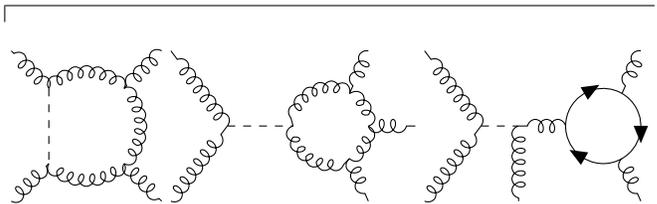
\begin{figure}
	\centering
	\begin{tikzpicture}[baseline={(0,0)},scale=0.5]
		\begin{feynman}
			\vertex (g1) at (-3,-1.5);
			\vertex (g2) at (-3,1.5);
			\vertex (g3) at (3,1.5);
			\vertex (g4) at (3,-1.5);
			\vertex (g5) at (-1,1.5);
			\vertex (g6) at (1,1.5);
			\vertex (v1) at (-2,0);
			\vertex (v3) at (-1,0);
			\vertex (v4) at (1,0);
			\vertex (v2) at (2,0);
			\diagram*{
				(g1) -- [gluon] (v1) -- [gluon] (g2),
				(g3) -- [gluon] (v2) -- [gluon] (g4),
				(v1) -- [scalar] (v3),
				(v3) -- [gluon, half right] (v4),
				(v4) -- [scalar] (v2),
				(g5) -- [gluon] (v3),
				(v4) -- [gluon] (g6),
			};
		\end{feynman}
	\end{tikzpicture}
	\caption{An example of a Feynman diagram that contributes to the tree-level amplitude with two axion propagators $\mathcal{A}^{(0)}_{\text{ax}^2}$.}
	\label{fig:treeax}
\end{figure}
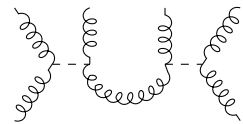

In twistorial theories, the QCD amplitude $\mathcal{A}_\text{QCD}^{(2)}$ contributes to $\mathcal{A}_\text{QCD+ax}^{(2)}$ with specific values of $n_f$. The QCD amplitude can be decomposed into three gauge invariant parts corresponding to different powers of $n_f$, as in eq.~\eqref{QCDdecomp}. Having three twistorial theories should then allow us to solve for these three amplitudes that comprise the full QCD amplitude. Restricting to $N_c=2$ or $3$, we get three equations for the three classes of twistorial theories appearing in tab.~\ref{tab:theories}:
\begin{align}
	\label{ampI}
    \begin{split}
    \mathcal{A}^{(2)}_\text{I}
	=&~
	\mathcal{A}^{(2)}_{G}
	+(N_c+6)
	\left(
	\mathcal{A}^{[g\to ax]}_\text{ax}
	+\mathcal{A}^{[g]}_\text{ax}
	\right)
    \\
	&+(N_c+6)^2\,\mathcal{A}^{(0)}_{\text{ax}^2}
    \end{split}
	\\
	\label{ampII}
	\mathcal{A}^{(2)}_\text{II}
	=&~
	\mathcal{A}^{(2)}_G
	+(N_c+6)\,\mathcal{A}^{(2)}_F
	+(N_c+6)^2\,\mathcal{A}^{(2)}_{F^2}
	\\
	\label{ampIII}
	\begin{split}
		\mathcal{A}^{(2)}_\text{III}
		=&~
		\mathcal{A}^{(2)}_G
		+N_c\,\mathcal{A}^{(2)}_F
		+N_c^2\,\mathcal{A}^{(2)}_{F^2}
		\\
		&+6\left(
		\mathcal{A}^{[g\to ax]}_\text{ax}
		+\mathcal{A}^{[g]}_\text{ax}
		+N_c\,\mathcal{A}^{[f]}_\text{ax}
		\right)
        \\
		&+6^2\mathcal{A}^{(0)}_{\text{ax}^2}
		\,.
	\end{split}
\end{align}
The solution to this system of equations is
\begin{align}
	\label{AGeq}
    \begin{split}
    \mathcal{A}^{(2)}_{G}
	=&~
	\mathcal{A}^{(2)}_\text{I}
	-(N_c+6)
	\left(
	\mathcal{A}^{[g\to ax]}_\text{ax}
	+\mathcal{A}^{[g]}_\text{ax}
	\right)
    \\
	&-(N_c+6)^2\,\mathcal{A}^{(0)}_{\text{ax}^2}
    \end{split}
	\\
	\label{AF1eq}
	\begin{split}
		\mathcal{A}^{(2)}_F
		=&~
		-\frac{2N_c+6}{N_c(N_c+6)}\,\mathcal{A}^{(2)}_\text{I}
		-\frac{N_c}{6(N_c+6)}\,\mathcal{A}^{(2)}_\text{II}
        \\
		&~+\frac{N_c+6}{6N_c}\,\mathcal{A}^{(2)}_\text{III}
		+\mathcal{A}^{[g\to ax]}_\text{ax}
		+\mathcal{A}_\text{ax}^{[g]}
		\\
        &~-(N_c+6)\,\mathcal{A}_\text{ax}^{[f]}
		+2(N_c+6)\,\mathcal{A}^{(0)}_{\text{ax}^2}
	\end{split}
	\\
	\label{AF2eq}
	\begin{split}
		\mathcal{A}^{(2)}_{F^2}
		=&~
		\frac{1}{N_c(N_c+6)}\,\mathcal{A}^{(2)}_\text{I}
		+\frac{1}{6(N_c+6)}\,\mathcal{A}^{(2)}_\text{II}
		\\
        &~-\frac{1}{6N_c}\,\mathcal{A}^{(2)}_\text{III}
		+\mathcal{A}_\text{ax}^{[f]}-\mathcal{A}^{(0)}_{\text{ax}^2}
		\,,
	\end{split}
\end{align}
where, again, the expressions are only valid for $N_c=2,3$.

The above solution expresses the two-loop QCD amplitude with arbitrary $n_f$ \eqref{QCDdecomp} in terms of two-loop amplitudes of the three twistorial theories, one-loop amplitudes involving a single axion propagator, and a tree-level amplitude with two internal axion propagators. The computational benefit is that $\mathcal{A}^{(2)}_\text{I}$, $\mathcal{A}^{(2)}_\text{II}$, and $\mathcal{A}^{(2)}_\text{III}$ can be computed using the CA bootstrap when restricted to the all-plus-helicity configuration. However, we expect that the CA bootstrap only gives the finite remainder of these twistorial amplitudes when computed with dimensional regularization, as was found for the 4-point amplitude in twistorial theory II \cite{Dixon:2024mzh}. So, eqs.~\eqref{AGeq}--\eqref{AF2eq} become equalities at the level of the finite remainders, and the divergent terms are determined by Catani's formula \cite{Catani:1998bh}. In the supplemental materials, we review what is known and what remains to be computed of the amplitudes appearing on the right-hand sides of eqs.~\eqref{AGeq}--\eqref{AF2eq}. To summarize, all single-trace terms are either known or conjectured, while the rational parts to the double- and triple-trace terms remain undetermined.

Finally, we emphasize that eqs.~\eqref{AGeq}--\eqref{AF2eq} hold for generic helicities. Extending the CA bootstrap to more generic helicity configurations is under active investigation.

\section{\label{sec:conclusions}Conclusions}

The CA bootstrap computes form factors of twistorial theories. Form factors with insertions of $\tfrac{1}{2}\tr(B^2)$ give helicity amplitudes of the full, non-self-dual gauge theory in the zero-operator-momenta limit, but these theories still contain matter content in their spectra that does not describe the interactions of QCD. We have nonetheless used the CA bootstrap to compute parts of ordinary two-loop QCD amplitudes. In particular, we have 1) computed the partial amplitude $A_n^{(-1,1);\,\text{all-plus}}$, valid for arbitrary $N_c$, by using supersymmetry Ward identities to relate it to a partial amplitude of twistorial theory III, which was computed using the CA bootstrap in ref.~\cite{Costello:2023vyy}, and 2) shown that the full-color all-plus-helicity $n$-gluon amplitude in QCD with an arbitrary number of quark flavors $n_f$ is given by CA bootstrap results and one-loop- and tree-level amplitudes involving the quartic axion, when $N_c=2$ or $3$. This latter result allows for these two-loop amplitudes to be computed using computationally tractable techniques like generalized unitarity \cite{Bern:1994zx, Bern:1994cg} and on-shell recursion \cite{Britto:2004ap,Britto:2005fq, Cachazo:2004kj, Bern:2005ji, Berger:2006ci}, bringing an $n$-gluon result for this helicity amplitude in reach \footnote{This reduction to lower-loop techniques for the two-loop all-plus amplitude agrees with the claims of refs.~\cite{Dunbar:2016aux,Dunbar:2016cxp,Dunbar:2016gjb,Dunbar:2017nfy,Dunbar:2019fcq,Dunbar:2020wdh,Dalgleish:2020mof,Dalgleish:2024sey,Kosower:2022bfv,Kosower:2022iju}}. These results are invaluable for understanding the analytic structure of QCD amplitudes and for making precise predictions for observables at hadron colliders.

\begin{acknowledgments}
    The author would like to thank Kevin Costello, Federica Devoto, Lance Dixon, and Natalie Paquette for helpful discussions, and he thanks Kevin Costello and Natalie Paquette for useful comments and suggestions on the draft. This work was supported in part by the US Department of Energy under contract DE--AC02--76SF00515 and by the Simons Collaboration on Celestial Holography.
\end{acknowledgments}

\bibliography{QCDfromCA}
 \bibliographystyle{apsrev4-1}
\clearpage

\onecolumngrid

\pagebreak

\widetext
\clearpage
\begin{center}
\textbf{\large Supplemental Materials}
\end{center}

\setcounter{equation}{0}
\setcounter{figure}{0}
\setcounter{table}{0}
\setcounter{section}{0}
\setcounter{page}{1}
\makeatletter
\renewcommand{\theequation}{S\arabic{equation}}
\renewcommand{\thefigure}{S\arabic{figure}}
\renewcommand{\thesection}{S\arabic{section}}

\section{Diagrams contributing to $A_n^{(-1,1)}$}
\label{sec:diagrams}
In this section, we show that $A_n^{(-1,1)}$ is solely built from planar Feynman diagrams with the topology $P_3$ shown in fig.~\ref{fig:tops}, i.e.~it does not receive any non-planar contributions. For $n=4$, it was shown that this statement is true for the all-plus-helicity configuration in ref.~\cite{Bern:2002zk}, and we have checked that it is true for $n=5$ with generic helicities by generating the explicit Feynman diagrams using QGRAF \cite{Nogueira:1991ex}. This contradicts a claim made in ref.~\cite{DeLaurentis:2023nss} that $A_5^{(-1,1)}$ receives contributions from non-planar diagrams.

\begin{figure}[h]
    \centering
    \begin{equation*}
    \begin{tikzpicture}[baseline={(0,0)},scale=0.75]
        \begin{feynman}
            \vertex (v1) at (-2,-1);
            \vertex (v2) at (-2,1);
            \vertex (v3) at (2,1);
            \vertex (v4) at (2,-1);
            \vertex (v5) at (0,1);
            \vertex (v6) at (0,0.5);
            \vertex (v7) at (0,-0.5);
            \vertex (v8) at (0,-1);
            \vertex (g1) at (-3,-2) {\(a_i\)};
            \vertex (g2) at (-3,2) {\(a_j\)};
            \vertex (g3) at (3,2) {\(a_k\)};
            \vertex (g4) at (3,-2) {\(a_l\)};
            \vertex (g5) at (1,0.5) {\(a_m\)};
            \vertex (g6) at (1,-0.5) {\(a_n\)};
            \vertex (a1) at (-2.75,0.25) {\(\vdots\)};
            \vertex (a2) at (2.75,0.25) {\(\vdots\)};
            \vertex (a3) at (0.75,0.2) {\(\vdots\)};
            \diagram*{
                (v1) -- [anti fermion] (v8)
                -- [anti fermion] (v4) 
                -- [anti fermion] (v3) 
                -- [anti fermion, swap] (v5)
                -- [anti fermion, swap] (v2) 
                -- [anti fermion] (v1),
                (v8) -- [gluon] (v7)
                -- [gluon] (v6) -- [gluon] (v5),
                (g1) -- [gluon] (v1),
                (v2) -- [gluon] (g2),
                (g3) -- [gluon] (v3),
                (v4) -- [gluon] (g4),
                (v6) -- [gluon] (g5),
                (v7) -- [gluon] (g6),
            };
        \end{feynman}
    \end{tikzpicture}
    =
    \tr(t^{a_i}\cdots t^{a_j}t^{b_1}t^{a_k}\cdots t^{a_l}t^{b_2})
    \,if^{b_1a_mc_1}\,if^{c_1a_{m+1}c_2}\,\cdots\,if^{c_{n-1}a_nb_2}
    \end{equation*}
    \caption{Non-planar color diagrams contributing to $\mathcal{A}^{(2)}_F$ are of the form shown. The dots represent other external legs.}
    \label{fig:nonplanar}
\end{figure}

The color factor(s) of any non-planar Feynman diagram can be decomposed into a linear combination of color factors of the form shown in fig.~\ref{fig:nonplanar}, where each term is related to each other by a permutation of the external legs denoted by $a_i$. The rules for reading color diagrams are
\begin{align}
\label{diagramrules}
\begin{split}
    \begin{tikzpicture}[baseline={(0,0)},scale=0.75]
        \begin{feynman}
            \vertex (g1) at (0,1.5) {\(a\)};
            \vertex (g2) at (1.5*1.7321/2,-1.5/2) {\(b\)};
            \vertex (g3) at (-1.5*1.7321/2,-1.5/2) {\(c\)};
            \vertex (v1) at (0,0);
            \diagram*{
                (g1) -- [gluon] (v1),
                (g2) -- [gluon] (v1),
                (g3) -- [gluon] (v1),
            };
        \end{feynman}
    \end{tikzpicture}
    &= if^{abc}
    \hspace{1cm}
    \begin{tikzpicture}[baseline={(0,0)},scale=0.75]
        \begin{feynman}
            \vertex (g1) at (-1.5,0) {\(a\)};
            \vertex (g2) at (1,0) {\(b\)};
            \diagram*{
                (g1) -- [gluon] (g2)
            };
        \end{feynman}
    \end{tikzpicture}
    = \delta^{ab}
    \\
    \begin{tikzpicture}[baseline={(0,0)},scale=0.75]
        \begin{feynman}
            \vertex (ti) at (-1.5*1.7321/2,0) {\(i\)};
            \vertex (tj) at (1.5*1.7321/2,0) {\(j\)};
            \vertex (v) at (0,0);
            \vertex (g) at (0,1.5) {\(a\)};
            \diagram*{
                (ti) -- [fermion] (v) 
                -- [fermion] (tj);
                (g) -- [gluon] (v)
            };
        \end{feynman}
    \end{tikzpicture}
    &= (t^a)^i_{~j}
    \hspace{0.95cm}
    \begin{tikzpicture}[baseline={(0,0)},scale=0.75]
        \begin{feynman}
            \vertex (ti) at (-1.5,0) {\(i\)};
            \vertex (tj) at (1,0) {\(j\)};
            \diagram*{
                (ti) -- [fermion] (tj)
            };
        \end{feynman}
    \end{tikzpicture}
    = \delta^i_{~j}
    \,.
\end{split}	
\end{align}
The decomposition of color factor(s) of Feynman diagrams into a permutation sum of terms of the form shown in fig.~\ref{fig:nonplanar} follows from using the commutation relation
\begin{align}
\label{commutator}
\begin{split}
    \begin{tikzpicture}[baseline={(0,0)},scale=0.75]
        \begin{feynman}
            \vertex (ti) at (-1.5*1.7321/2,0);
            \vertex (tj) at (1.5*1.7321/2,0);
            \vertex (v) at (0,0);
            \vertex (v1) at (0,1);
            \vertex (g1) at (-1,1.5) {\(a\)};
            \vertex (g2) at (1,1.5) {\(b\)};
            \diagram*{
                (ti) -- [fermion] (v) 
                -- [fermion] (tj);
                (v) -- [gluon] (v1);
                (g1) -- [gluon] (v1);
                (g2) -- [gluon] (v1);
            };
        \end{feynman}
    \end{tikzpicture}
    &=
    \begin{tikzpicture}[baseline={(0,0)},scale=0.75]
        \begin{feynman}
            \vertex (ti) at (-1.5*1.7321/2,0);
            \vertex (tj) at (1.5*1.7321/2,0);
            \vertex (v1) at (-0.5,0);
            \vertex (v2) at (0.5,0);
            \vertex (g1) at (-0.5,1.5) {\(a\)};
            \vertex (g2) at (0.5,1.5) {\(b\)};
            \diagram*{
                (ti) -- (v1) 
                -- [fermion] (v2) -- (tj),
                (g1) -- [gluon] (v1),
                (g2) -- [gluon] (v2),
            };
        \end{feynman}
    \end{tikzpicture}
    \;
    -
    \;
    \begin{tikzpicture}[baseline={(0,0)},scale=0.75]
        \begin{feynman}
            \vertex (ti) at (-1.5*1.7321/2,0);
            \vertex (tj) at (1.5*1.7321/2,0);
            \vertex (v1) at (-0.5,0);
            \vertex (v2) at (0.5,0);
            \vertex (g1) at (-0.5,1.5) {\(b\)};
            \vertex (g2) at (0.5,1.5) {\(a\)};
            \diagram*{
                (ti) -- (v1) 
                -- [fermion] (v2) -- (tj),
                (g1) -- [gluon] (v1),
                (g2) -- [gluon] (v2),
            };
        \end{feynman}
    \end{tikzpicture}
    \\
    if^{abc}\,t^c\hspace{0.5cm}&=\hspace{0.7cm}t^a\,t^b 
    \hspace{0.75cm} -
    \hspace{0.8cm} t^b\,t^a
    \,,
\end{split}
\end{align}
or by applying the Jacobi identity to contractions of structure constants
\begin{align}
\label{jacobi}
\begin{split}
    \begin{tikzpicture}[baseline={(0,0.25)},scale=0.75]
        \begin{feynman}
            \vertex (g3) at (1,-0.5) {\(c\)};
            \vertex (g4) at (-1,-0.5) {\(d\)};
            \vertex (v) at (0,0);
            \vertex (v1) at (0,1);
            \vertex (g1) at (-1,1.5) {\(a\)};
            \vertex (g2) at (1,1.5) {\(b\)};
            \diagram*[edges=gluon]{
                (g3) -- (v);
                (g4) -- (v);
                (v) -- (v1);
                (g1) -- (v1);
                (g2) -- (v1);
            };
        \end{feynman}
    \end{tikzpicture}
    &=
    \begin{tikzpicture}[baseline={(0,0)},scale=0.75]
        \begin{feynman}
            \vertex (g3) at (-1.5,-1) {\(d\)};
            \vertex (g4) at (1.5,-1) {\(c\)};
            \vertex (v1) at (-0.5,0);
            \vertex (v2) at (0.5,0);
            \vertex (g1) at (-1.5,1) {\(a\)};
            \vertex (g2) at (1.5,1) {\(b\)};
            \diagram*[edges=gluon]{
                (g3) -- (v1) -- (v2), 
                (g4) -- (v2),
                (g1) -- (v1),
                (g2) -- (v2),
            };
        \end{feynman}
    \end{tikzpicture}
    \;
    -
    \;
    \begin{tikzpicture}[baseline={(0,0)},scale=0.75]
        \begin{feynman}
            \vertex (g3) at (-1.5,-1) {\(d\)};
            \vertex (g4) at (1.5,-1) {\(c\)};
            \vertex (v1) at (-0.5,0);
            \vertex (v2) at (0.5,0);
            \vertex (g1) at (-1.5,1) {\(b\)};
            \vertex (g2) at (1.5,1) {\(a\)};
            \diagram*[edges=gluon]{
                (g3) -- (v1) -- (v2), 
                (g4) -- (v2),
                (g1) -- (v1),
                (g2) -- (v2),
            };
        \end{feynman}
    \end{tikzpicture}
    \\
    if^{abe}\,if^{ecd}\hspace{0.15cm}&=\hspace{0.5cm}
    if^{dae}\,if^{ebc} 
    \hspace{0.7cm} -
    \hspace{0.6cm} if^{dbe}\,if^{eac}
    \,.
\end{split}
\end{align}

\begin{figure}[h]
    \centering
    \begin{equation*}
    \begin{tikzpicture}[baseline={(0,0)},scale=0.75]
        \begin{feynman}
            \vertex (v1) at (-2,-1);
            \vertex (v2) at (-2,1);
            \vertex (v3) at (2,1);
            \vertex (v4) at (2,-1);
            \vertex (v5) at (0,1);
            \vertex (v6) at (0,0);
            \vertex (v7) at (0,-1);
            \vertex (v8) at (-1.5,-1);
            \vertex (v9) at (-0.3,-1);
            \vertex (v10) at (0.5,-1);
            \vertex (v11) at (1.7,-1);
            \vertex (g1) at (-3,-2) {\(a_i\)};
            \vertex (g2) at (-3,2) {\(a_j\)};
            \vertex (g3) at (3,2) {\(a_k\)};
            \vertex (g4) at (3,-2) {\(a_l\)};
            \vertex (g5) at (1,0) {\(a_m\)};
            \vertex (g6) at (-1.5,-2);
            \vertex (g7) at (-0.3,-2);
            \vertex (g8) at (0.5,-2);
            \vertex (g9) at (1.7,-2);
            \vertex (a1) at (-2.75,0.25) {\(\vdots\)};
            \vertex (a2) at (2.75,0.25) {\(\vdots\)};
            \vertex (a3) at (-1,-1.5) {\(\dots\)};
            \vertex (a4) at (-0.9,-2.3) {\(\underbrace{\hspace{1.2cm}}_L\)};
            \vertex (a3) at (1,-1.5) {\(\dots\)};
            \vertex (a4) at (1.1,-2.3) {\(\underbrace{\hspace{1.2cm}}_R\)};
            \diagram*{
                (v1) -- [anti fermion] (v7)
                -- [anti fermion] (v4) 
                -- [anti fermion] (v3) 
                -- [anti fermion, swap] (v5)
                -- [anti fermion, swap] (v2) 
                -- [anti fermion] (v1),
                (v7) -- [gluon] (v6) -- [gluon] (v5),
                (g1) -- [gluon] (v1),
                (v2) -- [gluon] (g2),
                (g3) -- [gluon] (v3),
                (v4) -- [gluon] (g4),
                (v6) -- [gluon] (g5),
                (v8) -- [gluon] (g6),
                (v9) -- [gluon] (g7),
                (v10) -- [gluon] (g8),
                (v11) -- [gluon] (g9),
            };
        \end{feynman}
    \end{tikzpicture}
    =
    \begin{tikzpicture}[baseline={(0,0)},scale=0.75]
        \begin{feynman}
            \vertex (v1) at (-2,-1);
            \vertex (v2) at (-2,1);
            \vertex (v3) at (3,1);
            \vertex (v4) at (3,-1);
            \vertex (v5) at (0,1);
            \vertex (v6) at (0.6,-1);
            \vertex (v7) at (0,-1);
            \vertex (v8) at (-1.5,-1);
            \vertex (v9) at (-0.3,-1);
            \vertex (v10) at (1.5,-1);
            \vertex (v11) at (2.7,-1);
            \vertex (g1) at (-3,-2) {\(a_i\)};
            \vertex (g2) at (-3,2) {\(a_j\)};
            \vertex (g3) at (4,2) {\(a_k\)};
            \vertex (g4) at (4,-2) {\(a_l\)};
            \vertex (g5) at (0.6,-2) {\(a_m\)};
            \vertex (g6) at (-1.5,-2);
            \vertex (g7) at (-0.3,-2);
            \vertex (g8) at (1.5,-2);
            \vertex (g9) at (2.7,-2);
            \vertex (a1) at (-2.75,0.25) {\(\vdots\)};
            \vertex (a2) at (3.75,0.25) {\(\vdots\)};
            \vertex (a3) at (-1,-1.5) {\(\dots\)};
            \vertex (a4) at (-0.9,-2.3) {\(\underbrace{\hspace{1.2cm}}_L\)};
            \vertex (a3) at (2,-1.5) {\(\dots\)};
            \vertex (a4) at (2.1,-2.3) {\(\underbrace{\hspace{1.2cm}}_R\)};
            \diagram*{
                (v1) -- [anti fermion] (v7)
                -- [anti fermion] (v4) 
                -- [anti fermion] (v3) 
                -- [anti fermion, swap] (v5)
                -- [anti fermion, swap] (v2) 
                -- [anti fermion] (v1),
                (v7) -- [gluon] (v5),
                (g1) -- [gluon] (v1),
                (v2) -- [gluon] (g2),
                (g3) -- [gluon] (v3),
                (v4) -- [gluon] (g4),
                (v6) -- [gluon] (g5),
                (v8) -- [gluon] (g6),
                (v9) -- [gluon] (g7),
                (v10) -- [gluon] (g8),
                (v11) -- [gluon] (g9),
            };
        \end{feynman}
    \end{tikzpicture}
    -
    \begin{tikzpicture}[baseline={(0,0)},scale=0.75]
        \begin{feynman}
            \vertex (v1) at (-3,-1);
            \vertex (v2) at (-3,1);
            \vertex (v3) at (2,1);
            \vertex (v4) at (2,-1);
            \vertex (v5) at (0,1);
            \vertex (v6) at (-0.6,-1);
            \vertex (v7) at (0,-1);
            \vertex (v8) at (-2.5,-1);
            \vertex (v9) at (-1.3,-1);
            \vertex (v10) at (0.5,-1);
            \vertex (v11) at (1.7,-1);
            \vertex (g1) at (-4,-2) {\(a_i\)};
            \vertex (g2) at (-4,2) {\(a_j\)};
            \vertex (g3) at (3,2) {\(a_k\)};
            \vertex (g4) at (3,-2) {\(a_l\)};
            \vertex (g5) at (-0.6,-2) {\(a_m\)};
            \vertex (g6) at (-2.5,-2);
            \vertex (g7) at (-1.3,-2);
            \vertex (g8) at (0.5,-2);
            \vertex (g9) at (1.7,-2);
            \vertex (a1) at (-3.75,0.25) {\(\vdots\)};
            \vertex (a2) at (2.75,0.25) {\(\vdots\)};
            \vertex (a3) at (-2,-1.5) {\(\dots\)};
            \vertex (a4) at (-1.9,-2.3) {\(\underbrace{\hspace{1.2cm}}_L\)};
            \vertex (a3) at (1,-1.5) {\(\dots\)};
            \vertex (a4) at (1.1,-2.3) {\(\underbrace{\hspace{1.2cm}}_R\)};
            \diagram*{
                (v1) -- [anti fermion] (v7)
                -- [anti fermion] (v4) 
                -- [anti fermion] (v3) 
                -- [anti fermion, swap] (v5)
                -- [anti fermion, swap] (v2) 
                -- [anti fermion] (v1),
                (v7) -- [gluon] (v5),
                (g1) -- [gluon] (v1),
                (v2) -- [gluon] (g2),
                (g3) -- [gluon] (v3),
                (v4) -- [gluon] (g4),
                (v6) -- [gluon] (g5),
                (v8) -- [gluon] (g6),
                (v9) -- [gluon] (g7),
                (v10) -- [gluon] (g8),
                (v11) -- [gluon] (g9),
            };
        \end{feynman}
    \end{tikzpicture}
    \end{equation*}
    \caption{The left side of the equation is a diagram representation of a term in the sum appearing in eq.~\eqref{nonplanto1leg}. The right side of the equation is the result of applying the commutation relation \eqref{commutator}.}
    \label{fig:nonplanar1leg}
\end{figure}
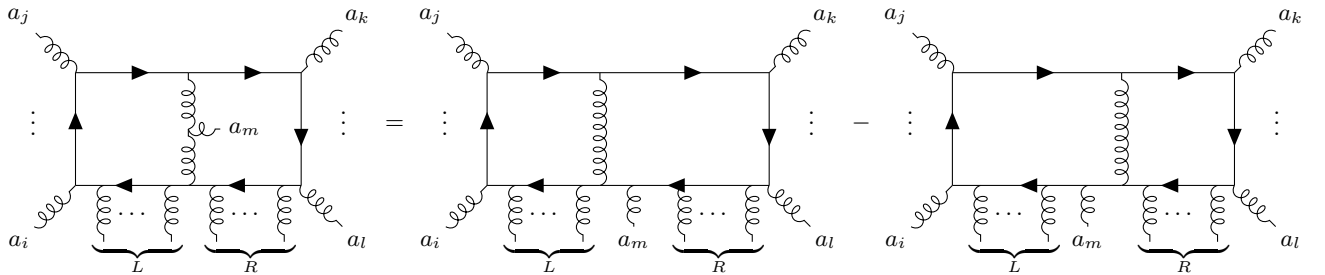

We will show that the diagram in fig.~\ref{fig:nonplanar} does not have a single-trace term, proving that $A_n^{(-1,1)}$ does not receive non-planar contributions. To see that the color diagram \ref{fig:nonplanar} has no single-trace terms, use the commutation relation \eqref{commutator} to express the diagram as a sum of diagrams of the form shown in fig.~\ref{fig:nonplanar1leg}. In symbols, 
\begin{multline}
    \label{nonplanto1leg}
    \tr(t^{a_i}\cdots t^{a_j}t^{b_1}t^{a_k}\cdots t^{a_l}t^{b_2})
    \,if^{b_1a_mc_1}\,if^{c_1a_{m+1}c_2}\,\cdots\,if^{c_{n-1}a_nb_2}
    \\
    =\sum_{R\subseteq M}(-1)^{|L|}\,
    \tr(t^{a_i}\cdots t^{a_j}t^{b_1}t^{a_k}\cdots t^{a_l}t^{a_R}t^{b_2}t^{a_L})\,
    if^{b_1a_mb_2}
    \,,
\end{multline}
where $M=(a_{m+1},a_{m+2},\dotsc,a_n)$ is an ordered list, $R$ is a sublist of $M$ ordered with respect to the ordering of $M$, and $L:=M\setminus R$ is also ordered with respect to $M$. We use the notation $t^{a_S}$ to denote the product
\begin{equation}
    t^{a_S}=t^{S_1}\,t^{S_2}\cdots t^{S_{|S|}}
\end{equation}
for some ordered list $S=(S_1,S_2,\dotsc,S_{|S|})$. For a given sublist $R$, we apply the commutation relation \eqref{commutator} and the $SU(N_c)$ Fierz identity
\begin{equation}
    (t^a)^i_j\,(t^a)^k_l = \delta^i_l\,\delta^k_j - \frac{1}{N_c}\delta^i_j\,\delta^k_l
\end{equation}
to get
\begin{multline}
    \tr(t^{a_i}\cdots t^{a_j}t^{b_1}t^{a_k}\cdots t^{a_l}t^{a_R}t^{b_2}t^{a_L})\,
    if^{b_1a_mb_2}
    \\
    =\tr(t^{a_i}\cdots t^{a_j}t^bt^{a_k}\cdots t^{a_l}t^{a_R}t^{a_m}t^bt^{a_L})
    -\tr(t^{a_i}\cdots t^{a_j}t^bt^{a_k}\cdots t^{a_l}t^{a_R}t^bt^{a_m}t^{a_L})
    \\
    =\tr(t^{a_k}\cdots t^{a_l}t^{a_R}t^{a_m})\,\tr(t^{a_L}t^{a_i}\cdots t^{a_j})
    -\frac{1}{N_c}\tr(t^{a_i}\cdots t^{a_l}t^{a_R}t^{a_m}t^{a_L})
    \\
    -
    \tr(t^{a_k}\cdots t^{a_l}t^{a_R})\,\tr(t^{a_m}t^{a_L}t^{a_i}\cdots t^{a_j})
    +\frac{1}{N_c}\tr(t^{a_i}\cdots t^{a_l}t^{a_R}t^{a_m}t^{a_L})
    \\
    =\tr(t^{a_k}\cdots t^{a_l}t^{a_R}t^{a_m})\,\tr(t^{a_L}t^{a_i}\cdots t^{a_j})
    -
    \tr(t^{a_k}\cdots t^{a_l}t^{a_R})\,\tr(t^{a_m}t^{a_L}t^{a_i}\cdots t^{a_j})
    \,,
\end{multline}
proving the claim. In fact, we have shown that two-loop non-planar diagrams involving a fermion loop only contribute to double-trace terms of $\mathcal{O}(N_c^0)$. Moreover, by taking the lists $(a_k,a_{k+1},\dotsc,a_l)$ and $R$ to both be empty, this argument shows that diagrams with topologies $P_1$ and $P_2$ do not produce $n_f/N_c$ single-trace terms. So, we can conclude that $A_n^{(-1,1)}$ only receives contributions from planar diagrams with topology $P_3$.

\section{Review of known results}
Let us review what is known and what still needs to be computed of the amplitudes on the right-hand-sides of eqs.~\eqref{AGeq}--\eqref{AF2eq}, when all helicities are positive. A summary is given in tab.~\ref{tab:knownamps}.
\begin{table}
    \centering
    \begin{tabular}{|c|c|c|c|c|c|c|c|}
    \hline
        Amplitudes & $\mathcal{A}^{(2)}_\text{I}$ & $\mathcal{A}^{(2)}_\text{II}$
		& $\mathcal{A}^{(2)}_\text{III}$ & $\mathcal{A}^{[g\to ax]}_\text{ax}$
		& $\mathcal{A}_\text{ax}^{[g]}$ & $\mathcal{A}_\text{ax}^{[f]}$
		& $\mathcal{A}^{(0)}_{\text{ax}^2}$ \\
        \hline
        Known & ST & Fully & ST & Fully (conj.) & CC & - & - \\
        Unknown & DT, TT & - & DT, TT & - & Rational & Fully & Fully \\
        \hline
    \end{tabular}
    \caption{A table summarizing the known and unknown parts of the amplitudes appearing on the right hand sides of eqs.~\eqref{AGeq}--\eqref{AF2eq}, when all gluons have positive helicities. By ``known'' we mean that an $n$-gluon expression has been computed; some of these amplitudes listed as unknown have known low-point results. The abbreviations ST, DT, TT, and CC mean ``single-trace,'' ``double-trace,'' ``triple-trace,'' and ``cut-constructable,'' respectively. The cut-constructable terms are those which can be completely determined from 4-dimensional unitarity cuts. The result for $\mathcal{A}_\text{ax}^{g\to ax}$ is given by a conjectured result in ref.~\cite{Dunbar:2020wdh}.}
    \label{tab:knownamps}
\end{table}

We begin with the twistorial amplitudes. The single-trace terms of $\mathcal{A}^{(2)}_\text{II}$ and $\mathcal{A}^{(2)}_\text{III}$ have been computed in ref.~\cite{Costello:2023vyy} using the CA bootstrap, and the double- and triple-trace terms of $\mathcal{A}^{(2)}_\text{II}$ were computed in ref.~\cite{Dixon:2024mzh}. The single-trace terms of $\mathcal{A}^{(2)}_\text{I}$ have not been explicitly written out, but they can be extracted from work done in ref.~\cite{Costello:2023vyy}. The result for the single-trace partial amplitude is
\begin{multline}
    A_{\text{I};n;1;1}^{(2)}(1,2,\dotsc,n)=
    \sum_{1\leq i<j<k<l\leq n}\frac{\spa{i}{j}\spa{j}{k}\spa{k}{l}\spa{l}{i}}{\spdenom{n}}
    \Bigg[
    -\frac{N_c^2+4}{2(4\pi)^4}
    \left(
    \frac{\spb{i}{j}\spb{k}{l}}{\spa{i}{j}\spa{k}{l}}
    +\frac{\spb{j}{k}\spb{l}{i}}{\spa{j}{k}\spa{l}{i}}
    \right)
    +
    \frac{4}{(4\pi)^2}\,\frac{\spb{i}{k}\spb{l}{j}}{\spa{i}{k}\spa{l}{j}}
    \\
    -\frac{4N_c}{(4\pi)^2}\,\frac{\spb{i}{j}\spb{k}{l}
    (\spa{i}{k}\spa{j}{l} + \spa{i}{l}\spa{j}{k})}{\spa{i}{j}^2\spa{k}{l}^2}
    +\frac{4N_c}{(4\pi)^2}\,\frac{\spb{i}{l}\spb{j}{k}
    (\spa{i}{k}\spa{l}{j} + \spa{i}{j}\spa{l}{k})}{\spa{i}{l}^2\spa{j}{k}^2}
    \Bigg]
    \,.
\end{multline}
The double- and triple-trace terms of $\mathcal{A}^{(2)}_\text{I}$ and $\mathcal{A}^{(2)}_\text{III}$ have not been published, but they will appear in a forthcoming paper by the author of this letter.
  
Parts of the one-loop all-plus amplitude with a single axion exchange $\mathcal{A}^{(1)}_\text{ax}$ are known and other parts can be obtained from a conjecture of ref.~\cite{Dunbar:2020wdh}. The single-trace part of this amplitude, $\mathcal{A}_\text{ax}^{[g\to ax]}$, is related to the conjecture of ref.~\cite{Dunbar:2020wdh}, which gives an $n$-point formula for the subleading-in-color partial amplitude $A_n^{(0,0);\text{all-plus}}$ \footnote{This conjecture has been analytically verified for $n=4,5,6,7$ external gluons using unitarity methods \cite{Bern:2000dn,Dalgleish:2020mof,Dunbar:2019fcq,Dalgleish:2024sey} and the rational terms have been numerically checked for $n=8,9$ \cite{Kosower:2022bfv,Kosower:2022iju}}. In particular, writing the single-trace axion amplitude in the trace basis as
\begin{equation}
    \mathcal{A}_\text{ax}^{[g\to ax]} = \sum_{S_n/\mathbb{Z}_n}\tr(t^{a_{\sig_1}}\cdots t^{a_{\sig_n}})\,A_n^{[g\to ax]}(\sig)
    \,,
\end{equation}
then 
\begin{align}
    \label{Aaxtran}
    A_n^{[g\to ax]}\big|_\text{tran} &= \frac{1}{6}\,A_n^{(0,0)}\big|_\text{tran}
    \\
    \label{Aaxrat}
    A_n^{[g\to ax]}\big|_\text{rat} &= \frac{1}{6}\,
    \left(
    A^{(0,0)}_{\text{III}; n} - A_n^{(0,0)}\big|_\text{rat} - A_n^{(-1,1)}
    \right)
    \,,
\end{align}
where we have split the partial amplitudes into the sum
\begin{equation}
    A = A\big|_\text{div} + A\big|_\text{tran} + A\big|_\text{rat}
    \,.
\end{equation}
The term $A\big|_\text{div}$ contains the IR and UV divergences in dimensional regularization as determined by Catani's formula after choosing a UV renormalization scheme \cite{Catani:1998bh}. The term $A\big|_\text{tran}$ consists of all the transcendental functions of the finite remainder, and $A\big|_\text{rat}$ is finite and a purely rational function in the kinematic variables. Eqs.~\eqref{Aaxtran} and \eqref{Aaxrat} follow from the fact that $A^{(0,0)}_{\text{III}; n}$ is finite, rational \cite{Costello:2023vyy}, and given by
\begin{equation}
    A^{(0,0)}_{\text{III}; n} = A_n^{(0,0)} + A_n^{(-1,1)} + 6\,A_n^{[g\to ax]}
    \,,
\end{equation}
where $A^{(0,0)}_{\text{III}; n}$ is the $\mathcal{O}(N_c^0)$ single-trace partial amplitude of the two-loop all-plus amplitude $\mathcal{A}^{(2)}_\text{III}$. It is easy to check eq.~\eqref{Aaxtran} for $n=4$. The amplitude $A_4^{[g\to ax]}$ only has triangle generalized-unitarity cuts, meaning that $A_4^{[g\to ax]}\big|_\text{tran}=0$, which agrees with known results \cite{Bern:2000dn,Bern:2002tk,Dunbar:2020wdh}. Though we do not do it here, it would be good to have an independent check of eq.~\eqref{Aaxtran} for $n=5$ and eq.~\eqref{Aaxrat} for $n=4,5$.

The remaining terms in the one-loop axion-exchange amplitude \eqref{1loopaxamp}, namely $\mathcal{A}^{[g]}_\text{ax}$ and $\mathcal{A}^{[f]}_\text{ax}$, are computed by sewing together the one-loop amplitude of positive-helicity gluons and one external off-shell axion, denoted $\mathcal{A}^{[J]}_\rho\equiv\mathcal{A}^{[J]}_\rho(p_\rho,1,\dotsc,n)$ with $J=g,f$, and its tree-level equivalent $\mathcal{A}^{(0)}_\rho\equiv\mathcal{A}^{(0)}_\rho(p_\rho,1,\dotsc,n)$, where the first argument $p_\rho$ is the off-shell momentum of the external axion. This sewing is the same procedure described in ref.~\cite{Dixon:2024tsb} to compute the one-loop all-plus amplitude from two copies of $\mathcal{A}^{(0)}_\rho$, except the permutation sum is slightly different. The sewing procedure is
\begin{equation}
	\mathcal{A}^{[J]}_\text{ax}(1,2,\dotsc,n) = 
	\sum_{c=3}^{n-1} \mathcal{A}^{[J]}_\rho(-P_{1,c-1},1,\dotsc,c-1)\frac{-i}{P_{1,c-1}^2P_{c,n}^2}
	\mathcal{A}^{(0)}_\rho(-P_{c,n},c,\dotsc,n)
	\;+\;
    \mathcal{P}\binom{n}{c-1}
	\,,
\end{equation}
where $+\,\mathcal{P}\binom{n}{c-1}$ instructs one to sum over all $\binom{n}{c-1}$ permutations of placing $c-1$ gluon momenta in the loop amplitude and $n-c+1$ momenta in the tree amplitude. This sewing of two amplitudes at an off-shell scalar leg is depicted in the middle and right diagrams of fig.~\ref{fig:loopax}.

The tree amplitude $\mathcal{A}^{(0)}_\rho$ was computed in ref.~\cite{Dixon:2004za} as the amplitude for gluon fusion into a Higgs boson in an effective field theory in which the top quark has infinite mass. The cut-constructible part of the one-loop amplitude with a gluon loop $\mathcal{A}^{[g]}_\rho$ is known for any number of external gluons \cite{Badger:2006us}. The rational part of $\mathcal{A}^{[g]}_\rho$ is only known up to $n=4$ gluons \cite{Badger:2006us,Berger:2006sh,Schmidt:1997wr}. The fermion-loop amplitude $\mathcal{A}^{[f]}_\rho$ has no four-dimensional unitarity cuts, so it is completely rational. It, too, is only known up to $n=4$ gluons \cite{Badger:2006us,Berger:2006sh,Schmidt:1997wr}. It should be possible to compute the rational terms using an augmented version of BCFW \cite{Britto:2004ap,Britto:2005fq}, similar to what was done for other one-loop amplitudes \cite{Bern:2005ji,Berger:2006sh}.

Finally, the remaining piece needed for the two-loop all-plus QCD amplitude as determined by eqs.~\eqref{AGeq}--\eqref{AF2eq} is the tree amplitude $\mathcal{A}^{(0)}_{\text{ax}^2}$, which has two internal axion propagators, as depicted in fig.~\ref{fig:treeax}. To compute this amplitude, one needs to sew together three amplitudes as follows
\begin{multline}
    \mathcal{A}^{(0)}_{\text{ax}^2}(1,2,\dotsc,n)
    =\sum_{b=3}^{n-3}\sum_{c=b+1}^{\floor{(n+b+1)/2}}
    \frac{2-\delta_{2c,n+b+1}}{2}\,\mathcal{A}^{(0)}_{2\rho}(P_{b,c-1},P_{c,n},1,\dotsc,b-1)
    \\
    \times
    \frac{-i}{P_{b,c-1}^4}\,\mathcal{A}^{(0)}_\rho(-P_{b,c-1},b,\dotsc,c-1)
    \,\frac{-i}{P_{c,n}^4}\,\mathcal{A}^{(0)}_\rho(-P_{c,n},c,\dotsc,n)
    \;+\; 
    \mathcal{P}\left\{\binom{n}{b-1}\times\binom{n-b+1}{c-b}\right\}
    \,,
\end{multline}
where $+\,\mathcal{P}\left\{\binom{n}{b-1}\times\binom{n-b+1}{c-b}\right\}$ instructs one to sum over all $\binom{n}{b-1}\times\binom{n-b+1}{c-b}$ permutations of placing $b-1$ momenta in $A_{2\rho}^{(0)}$, $c-b$ momenta in one copy of $A_{\rho}^{(0)}$, and the remaining $n-c+1$ momenta in the other copy of $A_{\rho}^{(0)}$. The factor of $(2-\delta_{2c,n+b+1})/2$ accounts for the fact that adding all of these permutations over counts by a factor of 2 when $2c=n+b+1$. The amplitude $\mathcal{A}^{(0)}_{2\rho}(p_{\rho_1},p_{\rho_2},1,\dotsc,n)$ is the tree-level amplitude with $n$ positive-helicity gluons and two external, off-shell axions. There is no known result for this amplitude.



\end{document}